\documentclass[pra,aps,onecolumn,superscriptaddress,showpacs,reprint]{revtex4}%
\usepackage{amsfonts}
\usepackage{amsmath}
\usepackage{amssymb}
\usepackage{graphicx}
\usepackage{epstopdf}%
\providecommand{\U}[1]{\protect\rule{.1in}{.1in}}

\begin{document}
\preprint{HEP/123-qed}
\title[ ]{Geometric entangling gates for coupled cavity system in decoherence-free subspaces}
\author{Yue-Yue Chen}
\affiliation{Laboratory of Photonic Information Technology, LQIT $\&$ SIPSE, South China
Normal University, Guangzhou 510006, China}
\author{Xun-Li Feng}
\affiliation{Laboratory of Photonic Information Technology, LQIT $\&$ SIPSE, South China
Normal University, Guangzhou 510006, China}
\affiliation{Department of Physics and Centre for Quantum Technologies, National University
of Singapore, 2 Science Drive 3, Singapore 117542}
\author{C.H. Oh}
\affiliation{Department of Physics and Centre for Quantum
Technologies, National University of Singapore, 2 Science Drive 3,
Singapore 117542} \keywords{one two three}

\pacs{ 03.67.Lx, 03.67.Pp, 03.65.Vf, 42.50.Pq }

\begin{abstract}
\textbf{Abstract.} We propose a scheme to implement geometric
entangling gates for two logical qubits in a coupled cavity system
in decoherence-free subspaces. Each logical qubit is encoded with
two atoms trapped in a single cavity and the geometric entangling
gates are achieved by cavity coupling and controlling the external
classical laser fields. Based on the coupled cavity system, the
scheme allows the scalability for quantum computing and relaxes the
requirement for individually addressing atoms.

\end{abstract}
\volumeyear{year}
\volumenumber{number}
\issuenumber{number}
\eid{identifier}
\startpage{1}
\endpage{5}
\maketitle

Exploiting appropriate coherent dynamics to generate entangling
gates between separate systems is of crucial importance to quantum
computing and quantum communication. Several schemes have been
proposed to engineer entangling gates \cite{SMB,chen_song,ZhengSB}
between atoms trapped in spatially separated cavities. It is
feasible and commonly used to mediate the distant optical cavities
by optical fiber \cite{J,S,SC}. However, decoherence resulted from
uncontrollable coupling to environment will collapse the state and
impair the performance for quantum process. Thus, decoherence is the
main obstacle for realizing quantum computing and quantum
information processing. In order to protect the fragile quantum
information and realize the promised speedup compared with classical
counterpart, a wealth of strategies have been proposed to deal with
decoherence. One efficient way is to construct a decoherence-free
subspace (DFS) if the interaction between quantum system and its
environment possesses some symmetry \cite{DFS}. Keeping a system
inside a DFS is regarded as a \textquotedblleft
passive\textquotedblright\ error-prevention approach while
error-correcting code, which is comprised of encoding information in
a redundant way, is regarded as an active approach \cite{Shor}.
Another promising strategy to cope with decoherence is based on the
mechanism of geometric phase \cite{Zanardi}. Geometric phases depend
only on some global geometric features of the evolution path and are
insensitive to local inaccuracies and fluctuations. However, the
total phases acquired during the evolution often consist of
geometric phases and the concomitant dynamic phases. Dynamic phases
may ruin the potential robustness of the scheme and should be
removed according to conventional wisdom. Literatures
\cite{convention} and \cite{LJ} proposed two simple methods to
remove dynamic phases. In contrast, the so-called unconventional
geometric gates, in which dynamic phases are not zero but
proportional to the geometric ones, were proposed \cite{WangXG,
shi}. The unconventional geometric gates were suggested to be
realized in cavity QED systems subsequently \cite{CQED1,CQED2}.

Schemes which combine the robust advantages of both DFS and the
geometric phase have been presented \cite{LX,X}. Reference \cite{LX}
exploits the spin-dependent laser-ion coupling in the presence of
Coulomb interactions, and then constructs a universal set of
unconventional geometric quantum gates in encoded subspaces.
Reference \cite{X} proposes to implement the geometric entangling
gates in DFS by using a dispersive atom-cavity interaction in a
single cavity. As is well known, the collective decoherence is often
regarded as a strict requirement for DFS strategy to overcome the
decoherence, however, such a requirement is largely relaxed in
\cite{X} because only two neighboring physical qubits, which encode
a logical qubit, are required to undergo collective dephasing. With
this merit, in this paper we extend the idea of \cite{X} to a
coupled cavity system where each cavity contains two atoms which
encode one logical qubit. In contrast to \cite{X}, the extension to
the coupled cavity system in this work allows the realization of
scalability of cavity QED based quantum computing by using the idea
of the distributed quantum computing \cite{JA} and relaxes the
requirement for individually addressing atoms.

Now let us describe our scheme more specifically. Considering two
coupled cavities which are linked with an optical fiber. We suppose
each cavity contains two $\Lambda$-type three-level atoms. For
convenience, we label the two cavities with $j$ and $k$,
respectively, the atoms in cavity $j$ ($k$) are denoted by
$j_{1},j_{2}\left(  k_{1},k_{2}\right)  $. The atomic level
configuration with couplings to the cavity modes and the driving
laser fields is shown in Fig. 1: $\left\vert e\right\rangle $ is an
excited state and $\left\vert 0\right\rangle $ and $\left\vert
1\right\rangle $ are two stable ground states, the latter two
constitute the basis of a physical qubit. Both transitions
$\left\vert 0\right\rangle \leftrightarrow\left\vert e\right\rangle
$ and $\left\vert 1\right\rangle \leftrightarrow$ $\left\vert
e\right\rangle $ are supposed to dispersively couple to the cavity
mode and be driven by two classical laser fields with opposite
detunings. One of the classical laser field acts on transitions
$\left\vert 0\right\rangle \leftrightarrow$ $\left\vert
e\right\rangle $ and $\left\vert 1\right\rangle
\leftrightarrow\left\vert e\right\rangle $ has a frequency $\omega$
closed to the cavity frequency $\omega_{c}$. Note that,
$\omega-\omega_{c}=\delta$, where $\delta$ is a small quantity. The
detuning of this classical field from the transition $\left\vert
m\right\rangle \leftrightarrow\left\vert e\right\rangle $ is
$\Delta_{m}=\omega_{m}-\omega$ $\left(  m=0,1\right)  $, where
$\omega_{m}$ is the energy difference between ground state
$\left\vert m\right\rangle $ and $\left\vert e\right\rangle $. The
corresponding detuning for the cavity mode is $\Delta_{m}+\delta$
(see Fig. 1). Similarly, the other laser with frequency
$\omega^{\prime}$ is tuned to satisfy the relation
$\omega_{m}-\omega^{\prime}=-\Delta_{m}$.

To overcome the collective dephasing, we encode the logical qubit in
the
cavity $j$ with a pair of physical qubits in a form $\left\vert 0_{j}%
\right\rangle ^{L}=\left\vert 0_{j_{1}}1_{j_{2}}\right\rangle $, $\left\vert
1_{j}\right\rangle ^{L}=\left\vert 1_{j_{1}}0_{j_{2}}\right\rangle $. The
subspace $\mathcal{C}_{j}^{2}=\left\{  \left\vert 0_{j}\right\rangle
^{L},\left\vert 1_{j}\right\rangle ^{L}\right\}  $ constitutes a DFS for the
single logical qubit $j$. Similarly, the logical qubit $k$ is encoded by the
two physical qubits $k_{1}$, $k_{2}$ in the cavity $k$.

\begin{figure}[ptb]
\begin{center}
\scalebox{0.8}{\includegraphics{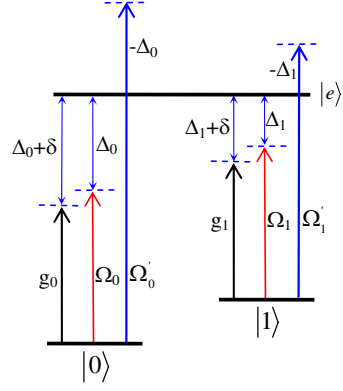}}
\end{center}
\caption{Atomic level structure and couplings. The transition $\left\vert
m\right\rangle (m=0,1) \leftrightarrow$ $\left\vert e\right\rangle $ is
coupled to the cavity mode with strength $g_{m}$ and driven by classical field
lasers with Rabi frequency $\Omega_{m}/\Omega_{m}^{\prime}$. }%
\label{fig1}%
\end{figure}

The coupling between the cavity fields and the fiber modes can be written as
the interaction Hamiltonian \cite{SMB}
\begin{equation}
H_{cf}=\underset{i=1}{\overset{\infty}{\sum}}\nu_{i}\left[  b_{i}\left(
a_{1}^{\dag}+\left(  -1\right)  e^{i\varphi}a_{2}^{\dag}\right)
+\text{H.c.}\right]  ,
\end{equation}
where $\nu_{i}$ is the coupling strength between fiber mode $i$ and the cavity
mode, $b_{i}$ is the annihilation operator for the fiber mode $i$ while
$a_{1}^{\dag}\left(  a_{2}^{\dag}\right)  $ is the creation operator for the
cavity mode $j$($k$), and $\varphi$ is the phase induced by the propagation of
the field through the fiber. In the short fiber limit, only resonant mode $b$
of the fiber interacts with the cavity mode. In this case, the Hamiltonian
$H_{cf}$ can be approximately written as \cite{SMB}%

\begin{equation}
H_{cf}=\nu\left[  b\left(  a_{1}^{\dag}+a_{2}^{\dag}\right)  +H.c.\right]  ,
\end{equation}
where the phase $\left(  -1\right)  e^{i\varphi}$ in $H_{cf}$ has been
absorbed into $a_{2}^{\dag}$ and $a_{2}$ \cite{chen_song}.

To implement the geometric entangling gate, we let the classical
laser fields plotted in Fig. 1 individually act on both atoms
$j_{1}$ and $k_{1}$. In the interaction picture, the Hamiltonian
describing the atom-field interaction takes the form
\begin{align}
H_{AC}  &  =\underset{l=j_{1},k_{1}}{\sum}\underset{m=0,1}{\sum}\frac
{\Omega_{m}^{\prime}}{2}e^{-i\Delta_{m}t}\left\vert e\right\rangle
_{l}\left\langle m\right\vert +\frac{\Omega_{m}}{2}e^{i\Delta_{m}t}\left\vert
e\right\rangle _{l}\left\langle m\right\vert +\underset{l=j_{1},j_{2}}{\sum}\underset{m=0,1}{\sum}g_{m}\left\vert
e\right\rangle _{l}\left\langle m\right\vert a_{1}e^{i\left(  \Delta
_{m}+\delta\right)  t}\nonumber\\
&  +\underset{l=k_{1},k_{2}}{\sum}\underset{m=0,1}{\sum}g_{m}\left\vert
e\right\rangle _{l}\left\langle m\right\vert a_{2}e^{i\left(  \Delta
_{m}+\delta\right)  t}+\text{H.c.}%
\end{align}

Following Ref. \cite{SMB}, we define three bosonic modes $c_{0}=\frac{1}{\sqrt{2}%
}\left(  a_{1}-a_{2}\right)  $, $c_{1}=\frac{1}{2}\left(  a_{1}+a_{2}+\sqrt
{2}b\right)  $, $c_{2}=\frac{1}{2}\left(  a_{1}+a_{2}-\sqrt{2}b\right)  $,
$c_{n}(n=0,1,2)$ are linearly relative to the field modes of the cavities and
fiber. Then we can rewrite the whole Hamiltonian in the interaction picture as%

\begin{equation}
H=H_{0}+H_{i},
\end{equation}
where%

\begin{equation}
H_{0}=\sqrt{2}\nu c_{1}^{\dag}c_{1}-\sqrt{2}\nu c_{2}^{\dag}c_{2},
\end{equation}
and%

\begin{align}
H_{i}  &  =\underset{m=0,1}{\underset{l=j_{1},k_{1}}{\sum}}\frac{\Omega
_{m}^{\prime}}{2}e^{-i\Delta_{m}t}\left\vert e\right\rangle _{l}\left\langle
m\right\vert +\frac{\Omega_{m}}{2}e^{i\Delta_{m}t}\left\vert e\right\rangle
_{l}\left\langle m\right\vert  +\underset{m=0,1}{\underset{l=j_{1},j_{2}}{\sum}}g_{m}\left\vert
e\right\rangle _{l}\left\langle m\right\vert \frac{1}{2}\left(  c_{1}%
+c_{2}+\sqrt{2}c_{0}\right)  e^{i\left(  \Delta_{m}+\delta\right)
t}\nonumber\\
&  +\underset{m=0,1}{\underset{l=k_{1},k_{2}}{\sum}}g_{m}\left\vert
e\right\rangle _{l}\left\langle m\right\vert \frac{1}{2}\left(  c_{1}%
+c_{2}-\sqrt{2}c_{0}\right)  e^{i\left(  \Delta_{m}+\delta\right)
t} +\text{H.c.}%
\end{align}
\ \ $\ \ \ \ $We now perform the unitary transformation $e^{iH_{0}t}$, and
obtain \cite{ZhengSB}

\begin{widetext}
\begin{align}
H_{i}=&\underset{m=0,1}{\underset{l=j_{1},k_{1}}{\sum}}\left(\frac{\Omega
_{m}^{\prime}}{2}e^{-i\Delta_{m}t}\left\vert e\right\rangle _{l}\left\langle
m\right\vert +\frac{\Omega_{m}}{2}e^{i\Delta_{m}t}\left\vert e\right\rangle
_{l}\left\langle m\right\vert\right)+ \underset{m=0,1}{\underset{l=j_{1},j_{2}}{\sum}}g_{m}\left\vert
e\right\rangle _{l}\left\langle m\right\vert \frac{1}{2}\left(  c_{1}%
e^{-i\sqrt{2}\nu t}+c_{2}e^{i\sqrt{2}\nu t}+\sqrt{2}c_{0}\right)  e^{i\left(
\Delta_{m}+\delta\right)  t}\nonumber\\
&+ \underset{m=0,1}{\underset{l=k_{1},k_{2}}{\sum}}g_{m}\left\vert
e\right\rangle _{l}\left\langle m\right\vert \frac{1}{2}\left(  c_{1}%
e^{-i\sqrt{2}\nu t}+c_{2}e^{i\sqrt{2}\nu t}-\sqrt{2}c_{0}\right)  e^{i\left(
\Delta_{m}+\delta\right)  t}+ \text{H.c.}%
\end{align}
\end{widetext}Here we assume that $\Delta_{m}\gg\sqrt{2}\nu,\delta,g_{m}$ and
$\Omega_{m}$ to make sure that atoms cannot exchange energy with the
fiber mode, cavity modes, and classical fields on account of the
large detuning. In this case, we may adiabatically eliminate the
excited atomic state considering no population transferred to the
this state. In order to cancel the Stark shifts caused by classical
laser fields, we set $\left\vert \Omega _{m}\right\vert =\left\vert
\Omega_{m}^{\prime}\right\vert $. Assuming further
$g_{m}\ll\Omega_{m}$, we can neglect the terms of $g_{m}^{2}$, which
indicate the Stark shifts caused by bosonic modes. From the above,
we obtain an effective Hamiltonian describing the coupling between
the atoms and bosonic modes assisted by the classical fields
\cite{JM}

\bigskip%
\begin{align}
H_{eff} &  =\left(  \left\vert 0\right\rangle _{j_{1}}\left\langle
0\right\vert +\left\vert 0\right\rangle _{k_{1}}\left\langle 0\right\vert
\right)  \lambda_{1}e^{-i\left(  \delta-\sqrt{2}\nu\right)  t}c_{1}^{\dag
}  +\left(  \left\vert 0\right\rangle _{j_{1}}\left\langle 0\right\vert
+\left\vert 0\right\rangle _{k_{1}}\left\langle 0\right\vert \right)
\lambda_{2}e^{-i\left(  \delta+\sqrt{2}\nu\right)  t}c_{2}^{\dag}\nonumber\\
&  +\left(  \left\vert 0\right\rangle _{j_{1}}\left\langle 0\right\vert
-\left\vert 0\right\rangle _{k_{1}}\left\langle 0\right\vert \right)
\lambda_{0}e^{-i\delta t}c_{0}^{\dag} +\left(  \left\vert 1\right\rangle _{j_{1}}\left\langle 1\right\vert
+\left\vert 1\right\rangle _{k_{1}}\left\langle 1\right\vert \right)
\lambda_{1}^{\prime}e^{-i\left(  \delta-\sqrt{2}\nu\right)  t}c_{1}^{\dag
}\nonumber\\
&  +\left(  \left\vert 1\right\rangle _{j_{1}}\left\langle 1\right\vert
+\left\vert 1\right\rangle _{k_{1}}\left\langle 1\right\vert \right)
\lambda_{2}^{\prime}e^{-i\left(  \delta+\sqrt{2}\nu\right)  t}c_{2}^{\dag
}  +\left(  \left\vert 1\right\rangle _{j_{1}}\left\langle 1\right\vert
-\left\vert 1\right\rangle _{k_{1}}\left\langle 1\right\vert \right)
\lambda_{0}^{\prime}e^{-i\delta t}c_{0}^{\dag}+\text{H.c.,}%
\end{align}
where

$\ \ \ \ \ \ \lambda_{0}=-\frac{\sqrt{2}\Omega_{0}g_{0}^{\ast}}{8}\left(
\frac{1}{\Delta_{0}}+\frac{1}{\Delta_{0}+\delta}\right)  $, $\ \ \ \ \ \ \lambda_{1}=-\frac{\Omega_{0}g_{0}^{\ast}}{8}\left(  \frac
{1}{\Delta_{0}}+\frac{1}{\Delta_{0}+\delta-\sqrt{2}\nu}\right)  $, $\ \ \ \ \ \ \lambda_{2}=-\frac{\Omega_{0}g_{0}^{\ast}}{8}\left(  \frac
{1}{\Delta_{0}}+\frac{1}{\Delta_{0}+\delta+\sqrt{2}\nu}\right)  $,

$\ \ \ \ \ \ \lambda_{0}^{\prime}=-\frac{\sqrt{2}\Omega_{1}g_{1}^{\ast}}%
{8}\left(  \frac{1}{\Delta_{1}}+\frac{1}{\Delta_{1}+\delta}\right)  $, $\ \ \ \ \ \ \lambda_{1}^{\prime}=-\frac{\Omega_{1}g_{1}^{\ast}}{8}\left(
\frac{1}{\Delta_{1}}+\frac{1}{\Delta_{1}+\delta-\sqrt{2}\nu}\right)  $, \ \ \ \ \ $\ \lambda_{2}=-\frac{\Omega_{1}g_{1}^{\ast}}{8}\left(  \frac
{1}{\Delta_{1}}+\frac{1}{\Delta_{1}+\delta+\sqrt{2}\nu}\right)  $.
\bigskip

Because the logical qubits $j$ and $k$ are located at different cavities, the
available DFS for the whole system is constructed by $\mathcal{C}_{jk}%
^{4}\equiv\mathcal{C}_{j}^{2}\otimes\mathcal{C}_{k}^{2}= \left\{  \left\vert
0_{j}^{L}0_{k}^{L}\right\rangle \text{, }\left\vert 0_{j}^{L}1_{k}%
^{L}\right\rangle \text{, }\left\vert 1_{j}^{L}0_{k}^{L}\right\rangle \text{,
}\left\vert 1_{j}^{L}1_{k}^{L}\right\rangle \right\}  ,$ and in this DFS the
Hamiltonian $H_{eff}$ is diagonal and takes the form%

\begin{equation}
H_{eff}= diag\left[  H_{0_{j}0_{k}}\text{, }H_{0_{j}1_{k}}\text{, }%
H_{1_{j}0_{k}}\text{, }H_{1_{j}1_{k}}\right]  ,
\end{equation}
where the diagonal matrix elements $H_{\mu_{j}\nu_{k}}(\mu,$ $\nu=0,1)$ are of
the form%

\begin{equation}
H_{\mu_{j}\nu_{k}}=\underset{n=0}{\overset{2}{\sum}}c_{n}^{\dag}\chi_{\mu
_{j}\nu_{k}}^{n}e^{-i\eta_{n}t}+\text{H.c.,}%
\end{equation}
where

$\ \ \ \chi_{0_{j}0_{k}}^{0}=0$, $\chi_{0_{j}0_{k}}^{1}=2\lambda_{1}$,
$\chi_{0_{j}0_{k}}^{2}=2\lambda_{2}$; $\ \ \ \chi_{0_{j}1_{k}}^{0}=\lambda_{0}-\lambda_{0}^{\prime}$, $\chi
_{0_{j}1_{k}}^{1}=\lambda_{1}+\lambda_{1}^{\prime}$, $\chi_{0_{j}1_{k}}%
^{2}=\lambda_{2}+\lambda_{2}^{\prime}$;

$\ \ \ \chi_{1_{j}0_{k}}^{0}=\lambda_{0}^{\prime}-\lambda_{0}$, $\chi
_{1_{j}0_{k}}^{1}=\lambda_{1}+\lambda_{1}^{\prime}$, $\chi_{1_{j}0_{k}}%
^{2}=\lambda_{2}+\lambda_{2}^{\prime}$; \ \ \ $\chi_{1_{j}1_{k}}^{0}=0$, $\chi_{1_{j}1_{k}}^{1}=2\lambda_{1}^{\prime}%
$, $\chi_{0_{j}0_{k}}^{2}=2\lambda_{2}^{\prime}$.\newline and

\ \ \ \ $\eta_{0}=\delta$, $\eta_{1}=\delta-\sqrt{2}\nu$, $\eta_{2}%
=\delta+\sqrt{2}\nu$.
\bigskip
Obviously, in the DFS $\mathcal{C}_{jk}^{4}$, time evolution matrix $U\left(
t\right)  $ also takes a diagonal form,%

\begin{equation}
U\left(  t\right)  =diag\left[  U_{0_{j}0_{k}}\text{, }U_{0_{j}1_{k}}\text{,
}U_{1_{j}0_{k}}\text{, }U_{1_{j}1_{k}}\right]  .
\end{equation}

\bigskip The corresponding diagonal matrix elements$\ U_{\mu_{j}\nu_{k}%
}\left(  t\right)  $ can be derived from Eq. (10) and they are in terms of
displacement operator%

\begin{align}
\ U_{\mu_{j}\nu_{k}}\left(  t\right)   &  =\hat{T}\exp\left[  -i\int_{0}%
^{t}H_{\mu_{j}\nu_{k}}\left(  \tau\right)  d\tau\right]  =\underset{n=0}{\overset{2}{\prod}}\exp\left(  {i\phi_{\mu_{j}\nu_{k}}^{n}%
}\right)  D\left(  \int_{c}d\alpha_{\mu_{j}\nu_{k}}^{n}\right)   =\exp\left[  {i}\phi_{\mu_{j}\nu_{k}}\right]  \underset{n=0}{\overset
{2}{\prod}}D\left(  \int_{c}d\alpha_{\mu_{j}\nu_{k}}^{n}\right)  \ ,
\end{align}

$\ $

with $\hat{T}$ being the time ordering operator, and%

\begin{equation}
\phi_{\mu_{j}\nu_{k}}=\overset{2}{\underset{n=0}{\sum}}\phi_{\mu_{j}\nu_{k}%
}^{n}=\overset{2}{\underset{n=0}{\sum}}\operatorname{Im}\left[  \int
_{c}\left(  \alpha_{\mu_{j}\nu_{k}}^{n}\right)  ^{\ast}d\alpha_{\mu_{j}\nu
_{k}}^{n}\right]  ,
\end{equation}

\begin{equation}
d\alpha_{\mu_{j}\nu_{k}}^{n}=-i\chi_{\mu_{j}\nu_{k}}^{n}e^{-i\eta_{n}\tau
}d\tau
\end{equation}

Considering the situation, where each bosonic mode is assumed initially in
vacuum state, the state of each bosonic mode evolves to coherent state at time
$t_{n}>0$. The corresponding amplitude $\int_{c}d\alpha_{\mu_{j}\nu_{k}}^{n}$
is dependent on the logic computational basis state $\left\vert \mu_{j}^{L}%
\nu_{k}^{L}\right\rangle $. It is not difficult to obtain $\alpha_{\mu_{j}%
\nu_{k}}^{n}$ by integrating Eq. (14)
\begin{equation}
\alpha_{\mu_{j}\nu_{k}}^{n}=\frac{\chi_{\mu_{j}\nu_{k}}^{n}}{\eta_{n}}\left(
e^{-i\eta_{n}t}-1\right)  .
\end{equation}

The above equation indicates that there is a time period $T$ fulfilling the
relation $T=2\pi l_{n}/\eta_{n}$, where $l_{n}$ is a positive integer and
$n=0,1,2$, in which the bosonic mode $c_{n}$ completes $l_{n}$ evolutions and
returns to its initial vacuum state. During this process the system
accumulates the following total phase%

\begin{align}
\gamma_{\mu_{j}\nu_{k}}(T)  &  =\phi_{\mu_{j}\nu_{k}}(T)  =-\overset{2}{\underset{n=0}{\sum}}\frac{2\pi l_{n}}{\eta_{n}}\left\vert
\chi_{\mu_{j}\nu_{k}}^{n}\right\vert ^{2}  =\gamma_{\mu_{j}\nu_{k}}^{g}+\gamma_{\mu_{j}\nu_{k}}^{d},
\end{align}
where $\gamma_{\mu_{j}\nu_{k}}^{d}$ and
$\gamma_{\mu_{j}\nu_{k}}^{g}$ stand for the dynamical and geometric
phases respectively, and can be calculated by
using the coherent state path integral method \cite{MH}%

\begin{align}
\gamma_{\mu_{j}\nu_{k}}^{d}  &  =\overset{2}{\underset{n=0}{\sum}}-\int
_{0}^{T}H_{\mu_{j}\nu_{k}}^{n}\left(  \left(  \alpha_{\mu_{j}\nu_{k}}%
^{n}\right)  ^{\ast},\alpha_{\mu_{j}\nu_{k}}^{n};t\right)  dt  =-\overset{2}{\underset{n=0}{\sum}}\frac{4\pi l_{n}}{\eta_{n}^{2}%
}\left\vert \chi_{\mu_{j}\nu_{k}}^{n}\right\vert ^{2},
\end{align}

\begin{equation}
\gamma_{\mu_{j}\nu_{k}}^{g}=\gamma_{\mu_{j}\nu_{k}}-\gamma_{\mu_{j}\nu_{k}%
}^{d}=\overset{2}{\underset{n=0}{\sum}}\frac{2\pi l_{n}}{\eta_{n}^{2}%
}\left\vert \chi_{\mu_{j}\nu_{k}}^{n}\right\vert ^{2},
\end{equation}
we find $\gamma_{\mu_{j}\nu_{k}}=-\gamma_{\mu_{j}\nu_{k}}^{g}=\frac{1}%
{2}\gamma_{\mu_{j}\nu_{k}}^{d}$. Thus the total phase $\gamma_{\mu_{j}\nu_{k}%
}$ and dynamical phase $\gamma_{\mu_{j}\nu_{k}}^{d}$ possess global geometric
features as does the geometric phase $\gamma_{\mu_{j}\nu_{k}}^{g}$. Therefore
at time $t=T=2\pi l_{n}/\eta_{n}$ the time evolution matrix takes the form%

\begin{equation}
U\left(  T\right)  =diag\left[  e^{i\gamma_{0_{j}0_{k}}}\text{, }%
e^{i\gamma_{0_{j}1_{k}}}\text{, }e^{i\gamma_{1_{j}0_{k}}}\text{, }%
e^{i\gamma_{1_{j}1_{k}}}\text{ }\right]  .
\end{equation}
$U\left(  T\right)  $ is actually the geometric entangling gate operation we
are targeting at and $U\left(  T\right)  $ is a nontrivial entangling gate
when the condition $\gamma_{0_{j}0_{k}}+\gamma_{1_{j}1_{k}}\neq\gamma
_{0_{j}1_{k}}+\gamma_{1_{j}1_{k}}$ is fulfilled \cite{X}.

We now give a brief discussion about the decoherence mechanisms of our scheme:
atomic spontaneous emission, cavity decay and fiber loss. Considering none of
the atoms are initially populated in the excited state since the quantum
information is encoded in ground states, and atoms cannot exchange energy with
the fiber mode, cavity modes and classical fields due to the large detuning,
thus no population is transferred to the excited atomic state. In this sense,
the spontaneous emission of the atomic excited state can be ignored.

Regarding the cavity decay and the fiber loss, the fidelity of the
resulting gates will be greatly impaired by them because the
geometric phases are acquired by the evolution of the optical modes.
So, strictly speaking, our scheme requires ideal good cavities and
fiber. However, if the mean number of photons of the optical fields
is sufficiently small, the cavities and fiber are normally not
excited and the moderate cavity decay and fiber loss can thus be
tolerated. For a coherent state the mean number of photons is equal
to the square of the amplitude of the state which is determined by
Eq.(15). Thus when the condition
$\frac{\chi_{\mu_{j}\nu_{k}}^{n}}{\eta_{n}}\ll1$ is fulfilled
\cite{CQED1}, the mean number of photons of the coherent state is an
even smaller number and can be regarded as a sufficiently small
number to ignore the effect of cavity and fiber decay. Now let us
use an example for further explanation. We choose the following
experimentally achievable parameters \cite {Kimble} $\nu/2\pi=26.72$
MHz, $g_{0}/2\pi=g_{1}/2\pi=20$ MHz, $\Omega_{0}/2\pi
=\Omega_{1}/2\pi=120$ MHz, $\Delta_{0}/2\pi=3000$ MHz,
$\Delta_{1}/2\pi=600$ MHz, $\delta/2\pi=35$ MHz. These parameters
satisfy the requirement $\frac{\chi_{\mu
_{j}\nu_{k}}^{n}}{\eta_{n}}\ll1$ and the approximation conditions
adopted in our derivation. The resulting entangling gate
corresponding to these parameters is $U(t)=diag\left\{
e^{0.1248i},e^{1.056i},e^{1.056i},e^{i\pi }\right\}  $ with the gate
operation time $t\approx0.3448$ $\mu$s. Obviously the gate operation
time is much shorter than the photon lifetime in optical cavities
\cite{AA}. According to Eq. (15) the amplitude of the coherent state
is dependent on the atomic states, for the above parameters the
amplitude corresponding to state $\left\vert
1_{j}^{L}1_{k}^{L}\right\rangle $ takes the maximal value, and the
maximal mean number of photons is 0.1087. In this case, the optical
modes are hardly excited and thus the moderate cavity decay and
fiber loss can be tolerated.

In conclusion, we have proposed a scheme to implement geometric
entangling gates for two logical qubits in a coupled cavity system
in DFS. Our scheme possesses both advantages of DFS and the
geometric phase. Besides, in comparison with the scheme of Ref.
\cite{X} which works in a single cavity, the scheme proposed in this
paper can easily realize the scalability of cavity QED-based quantum
computing by using the idea of the distributed quantum
computing\cite{JA} and can relax the requirement for individually
addressing atoms.

The work is supported by the NSFC under Grant No. 11074079, the
Ph.D. Programs Foundation of Ministry of Education of China, the
Open Fund of the State Key Laboratory of High Field Laser Physics (
Shanghai Institute of Optics and Fine Mechanics), and National
Research Foundation and Ministry of Education, Singapore, under
research Grant No. WBS: R-710-000-008-271.

\end{document}